\documentclass[twocolumn,showpacs,preprintnumbers,amsmath,amssymb]{revtex4}
\usepackage{amsfonts}
\usepackage{amsmath}
\usepackage{graphicx}
\usepackage{dcolumn}
\usepackage{bm}
\usepackage{overpic}
\usepackage{booktabs}

\newcommand{\PreserveBackslash}[1]{\let\temp=\\#1\let\\=\temp}
\newcolumntype{C}[1]{>{\PreserveBackslash\centering}p{#1}}
\newcolumntype{R}[1]{>{\PreserveBackslash\raggedleft}p{#1}}
\newcolumntype{L}[1]{>{\PreserveBackslash\raggedright}p{#1}}
\begin{document}

\title{Quantifying the influence of scientists and their publications: Distinguish prestige from popularity}
\author{Yan-Bo Zhou$^{1}$}
\author{Linyuan L\"{u}$^{1}$}\email{linyuan.lue@unifr.ch}
\author{Menghui Li$^{2,3}$}

\affiliation{$^1$Department of Physics, University of Fribourg, Chemin du Mus\'{e}e 3, CH-1700 Fribourg, Switzerland\\$^2$Temasek Laboratories, National University of Singapore, 117508, Singapore\\$^3$Beijing-Hong Kong-Singapore Joint Center for Nonlinear and Complex Systems (Singapore), National University of Singapore, Kent Ridge 119260, Singapore}

\date{\today}

\begin{abstract}
The number of citations is a widely used metric to evaluate the scientific credit of papers, scientists and journals. However, it does happen that a paper with fewer citations from prestigious scientists is of higher influence than papers with more citations. In this paper, we argue that from whom the paper is being cited is of higher significance than merely the number of received citations. Accordingly, we propose an interactive model on author-paper bipartite networks as well as an iterative algorithm to get better rankings for scientists and their publications. The main advantage of this method is twofold: (i) it is a parameter-free algorithm; (ii) it considers the relationship between the prestige of scientists and the quality of their publications. We conducted real experiments on publications in econophysics, and applied this method to evaluate the influences of related scientific journals. The comparisons between the rankings by our method and simple citation counts suggest that our method is effective to distinguish prestige from popularity.
\end{abstract}




\maketitle

\section{Introduction}
How to measure the scientific influence of scientists and their publications is a long-term debate. Although many metrics have been introduced, their objectiveness and effectiveness are always questioned \cite{Seglen1997,Favaloro2008,Maslov2008,Adler2009,Bruno2010}. Without a clear-cut criteria, nobody can tell whether it is fair enough to reflect the truth. It is well-known that the number of citations is the simplest indicator of scientific impact \cite{Garfield1955,Garfield1979,Amsterdamska1989}. Previous studies showed that the number of citations has certain correlation with the quality of research \cite{Trajtenberg1990,Aksnes2006,Moed2005}, which has thus been widely used to assess the scientific productions of individuals or institutions as well as the scientists' influence \cite{Van2005,Boyack2003,Mazloumian2011}. However, the value of each citation is indeed dependent on the $quoters$, i.e. the researchers who cited the paper \cite{Radicchi2009}.
If a paper is cited by prestigious scientists, it is probably a gem (i.e., high quality) and is thus highly appreciated. This important perspective is not considered in many citation-based ranking methods. Even the well-known $h$ index, defined as the number of papers with citation number $\geq h$, also treats each citation equally no matter who is the contributor \cite{Hirsch2005,Hirsch2007}.

In general, if a scientist cites a paper, it indicates that she endorses this paper as well as its authors. This can be considered as a spread of prestige (i.e., quality, will be quantified later) which cannot be reflected by the mere number of citations, since citation counts only reflect the popularity, but not the quality or prestige \cite{Franceschet2010}. As inspired by the success of Google's ranking system for web pages, the popular PageRank algorithm as well as some of its variants have been used to reflect the prestige in citation network of journals \cite{Bollen2006}, publications \cite{Chen2007,Walker2007,Ma2008} and scientists \cite{Radicchi2009,Ding2009}. Since the network analyzed is a particular projection of citation network, the result depends on how weights are assigned to links. In addition, the choice of damping factor in the PageRank-based methods also affects the results \cite{Ding2009}. It has been pointed out that different from the boredom attrition factor 0.15 of web surfers, the appropriate factor is 0.5 in the context of citations, corresponding to a citation chain to two links \cite{Chen2007}.

All previous studies focused on the ranking of either scientists or publications, while totally neglecting the fact that these two sides are interacting with each other. In other words, the scientists' prestige and quality of their publications are strongly correlated. It is obvious that a paper is expected to be of high quality if it was cited by prestigious scientists, meanwhile a high quality paper can raise prestige of its authors. From this perspective, we propose an iterative algorithm to quantify the quality of papers and scientists' prestige via considering their relationship on an author-paper bipartite network. The network is a directed bipartite network with two kinds of links. The link from author to paper represents the citing relationship while the link from paper to author indicates the authorship. Our method is parameter-free and can simultaneously obtain the ranking lists of papers and scientists. We perform our method on the dataset consists of 1990 scientists in the field of econophysics and their 2012 papers which are published between April 1995 and September 2010, and compare the results with citation counts (CC Rank). Although our method has overlap with CC rank, it also reveals significant and meaningful differences. The outliers indicate that some scientists or papers with low CC rank have higher influence than their citations indicate, while some are over evaluated by merely counting the number of citations.

\begin{table}[!htb]
\begin{center}
\caption{The journals (and an e-print server) that published more than five papers in our dataset.}
\begin{tabular}{rcc}
\hline
Journal &Abbreviation &Paper\\
\hline
Phsica A	&Physa	&1120\\
Phys Rev E	&PRE	&179\\
arXiv.org	&arXiv	&161\\
Eur Phys J B	&EPJB	&148\\
Quant Financ	&QF	&52\\
Int J Mod Phys C	&IJMPC	&47\\
Phys Lett A	&PLA	&31\\
Int J Theor Appl Financ	&IJTAF	&24\\
EPL-Europhys Lett	&EPL	&21\\
Phys Rev Lett	&PRL	&20\\
J Korean Phys Soc	&JKPS	&18\\
Adv Complex Syst	&ACS	&15\\
J Phys A-Math Theor	&JPA	&14\\
Proc Natl Acad Sci USA	&PNAS	&12\\
Acta Phys Pol B	&APPB	&11\\
J Stat Mech-Theory E	&JSM	&10\\
Chinese Phys Lett	&CPL	&7\\
Int J Mod Phys B	&IJMPB	&7\\
\hline\label{Journal}
\end{tabular}
\end{center}
\end{table}

\section{Data Description}
Our database consists of a set of papers in the field of
econophysics which are published between April 1995 and September 2010 in 78
scientific journals and an e-print server (i.e., arXiv.org). The data is obtained by filtering the whole set
of papers with keywords \emph{econophysics}, \emph{market},
\emph{finance}, \emph{stock}, \emph{price}, \emph{minority game},
\emph{money}, \emph{wealth}, \emph{trade} and \emph{GDP}. Finally,
we have 2012 papers and 1990 distinct authors. Actually, this data is an extension of the dataset analyzed in Refs. \cite{Fan2004,Li2005}. Among the 78 journals, more than half of them contain only one or two papers, and more than half of the papers are published in \emph{Physica A}. Table ~\ref{Journal} summaries the journals that published more than five papers in our dataset. The list of
references at the end of each paper is used to construct a paper
citation network. Note that only the papers within this dataset are
considered in the citation network. Thus the degree of a paper in
this citation network is indeed smaller than its actual number of citations according to the \emph{ISI Web of Science}. Unless otherwise stated,
\emph{citation} in our context always refers to the case within the paper citation
network.

\section{Construction of the Author-Paper Interactive Network}
\begin{figure}[h!]
\centering {\includegraphics[width=6.8cm]{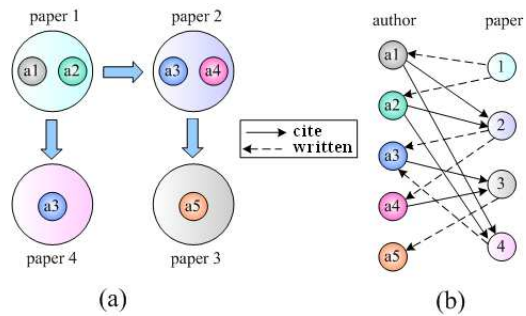}
}\caption{ (Color online) An illustration of how the author-paper bipartite network is constructed.
(a) The citation relations between four papers written by five scientists. (b) The
corresponding author-paper interactive network of (a), where the
directed links from an author to papers indicate that the author cites these
papers, and the directed links from a paper to authors mean that this paper is
written by these scientists. Here self-citations are not included.}\label{example}
\end{figure}

According to the paper citation relations and the authors of each
paper, we can obtain a directed author-paper interactive bipartite
network. The directed links from an author to papers denote that this author
cites these papers, while the directed links from a paper to authors represent
that the paper is co-authored by these scientists. Denote by $S$ and $P$ the
sets of scientists and papers, respectively, there are in total $M=|S|$ scientists
and $N=|P|$ papers. $A$ is an $M\times N$ adjacency matrix
representing the \emph{cite} relations between authors and papers, with
element $a_{i\alpha}=1$ if author $s_i$ cites paper $p_\alpha$,
and 0 otherwise. Similarly, $B$ is an $N\times M$ adjacency matrix
representing the \emph{written} relation, with element
$b_{\alpha i}=1$ if author $s_i$ is one of the authors of paper
$p_\alpha$, and 0 otherwise. Consider a paper $p_\alpha$ written by $n$
authors $s_1$, $s_2$,$\cdots$,$s_n$, citing paper $p_\beta$, then
there will be $n$ directed links from paper $p_\alpha$ respectively to
$n$ authors of $p_\alpha$, and $n$ directed links from $n$ authors
to paper $p_\beta$. Note that self-citations are not included in
the network. Figure~\ref{example} shows an illustration of how the author-paper bipartite network is constructed. In this example, there are five scientists and four papers with citation relation shown in
Fig.~\ref{example} (a). Paper $p_1$ cites papers $p_2$ and $p_4$,
paper $p_2$ cites paper $p_3$. Since $p_1$ is co-authored by
scientists $a_1$ and $a_2$, there are two links from $p_1$
respectively to $a_1$ and $a_2$ indicating that $a_1$ and $a_2$ are
the authors of paper $p_1$. According to the citation relation
between paper $p_1$ and $p_2$ (i.e., $p_1$ cites $p_2$), there are
two links respectively from $a_1$ and $a_2$ to $p_2$. By following the
same rules, we finally obtain the author-paper directed bipartite
network as shown in Fig.~\ref{example}(b).

\section{Ranking algorithm}
The advantage to use author-paper bipartite network is significant.
It utilizes the interactions between
reputation and publications of a scientist. Normally, a paper is expected to have high
quality if it was cited by prestigious scientists, while high
quality papers raise the scientists' prestige accordingly. Based on this
assumption, we defined an iterative algorithm on the author-paper bipartite network (AP Rank) to evaluate the impact
and prestige of papers and scientists. Our method can simultaneously
obtain the ranking lists of papers and scientists.

We denote $Q_{s_i}$ as the score of author $s_i$ to quantify $s_i$'s prestige, and $Q_{p_\alpha}$ the score of paper $p_\alpha$
to evaluate the quality of $p_\alpha$. For simplicity, we
consider the contribution of each author on a paper to be the same regardless of order. And
of course, one can assign weight on author's importance according
to her author rank in the paper. The score of a paper will
be evenly distributed to all its co-authors. This implies that if a
paper has more authors, each of them obtains less. Thus the score of
author $s_i$ is counted by summing over the scores that distributed
from all his papers. Mathematically, it reads
\begin{equation}
Q_{s_i}=\sum\limits_{\alpha \in P}\frac{Q_{p_\alpha}}{k_{p_\alpha}^{out}} \cdot b_{\alpha i},\label{PtoS}
\end{equation}
where $k_{p_\alpha}^{out}=\sum_{i\in S}b_{\alpha i}$ is the number of authors of paper $p_\alpha$. We will show later that $Q_{p_\alpha}$ should be a normalized score. This step is equivalent to mass diffusion from paper to author on the bipartite network, which is indeed a conservative process. We define a process to be conservative if the initial mass of the network is equal to the final mass after the process has taken place.

Unlike the diffusion process from paper to author, we adopt a non-conservative process from author to paper. We assume that if an author cites a paper, this means the author votes for (i.e., gives approval to the impact of this paper) this paper with score equals to his score $Q_{s}$. Clearly, if two papers have identical citation, the paper cited by prestigious scientists (i.e., authors with higher score) is more significant than the other papers. Here we assume that each paper inherently hold one score. By doing this, we are able to compare the performance of two authors who have zero citation according to their productivity. Accordingly, the score of paper $p_\alpha$ is equal to the summation of its inherent score and the total voting scores from the authors who cite it, namely
\begin{equation}
Q_{p_\alpha}=1+\sum\limits_{i \in S}{Q_{s_i} \cdot a_{i\alpha}}.
\end{equation}
To avoid the exponential increasing of the total score, we normalize the score in the following way,
\begin{equation}
\tilde{Q}_{p_\alpha}=Q_{p_\alpha}\cdot\frac{C}{\sum\limits_{\beta\in
P}Q_{p_\beta}},\label{StoP}
\end{equation}
where $C$ is the initial total score. At the beginning, we assign to each paper one unit of score. Thus $C=N$. Then the scores iterate following the link direction according to the above rules. We define the deviation of the two score vectors of paper between two iteration steps as
\begin{equation}
\Delta(t)=|\tilde{Q}_p(t)-\tilde{Q}_p(t-1)|=\frac{1}{N}\sum\limits_{\alpha\in P}{[\tilde{Q}_{p_\alpha}(t)-\tilde{Q}_{p_\alpha}(t-1)]^{2}}.
\end{equation}
The final scores are obtained when $\Delta(t)<\delta$, where $\delta$ represents the \emph{a priori} fixed precision. Here, we set $\delta = 10^{-4}$.

\section{Results}

\begin{figure}[h!]
\centering \scalebox{0.58}[0.58]{\includegraphics{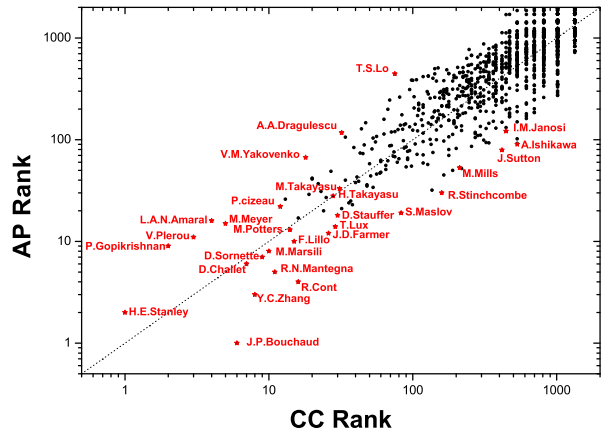}
}\caption{(Color online)Scatter plot of AP rank \emph{vs.} CC
rank for authors. If the two methods provide the same ranking all
the points would fall on the diagonal. The outliers indicate
the significant difference between AP rank and CC rank. We label some
typical examples in red. The Kendall's $\tau$ coefficient is 0.784.}\label{author}
\end{figure}

To test the above algorithm, we apply it to rank scientists and papers in the
field of econophysics. We have tested that the ranking results are
the same no matter the iteration start from the author side or from
the paper side. For an author, her citation
is the total number of citations received from other papers.
Similarly, for a paper, its citation is the number of papers that
cite this paper (i.e., the in-degree in the paper citation network).

\begin{figure*}[!htpb]
\centering \scalebox{1}[1]{\includegraphics{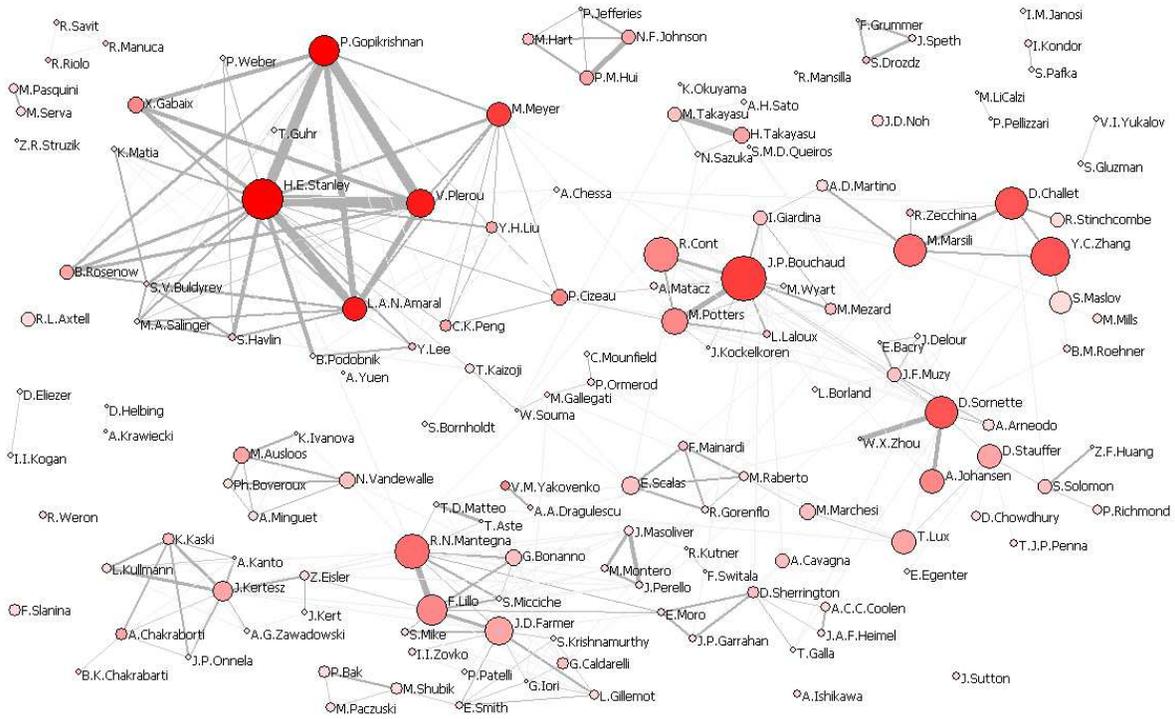}
}\caption{(Color online) The co-authorship network in the field of
econophysics. The top-150 scientists ranked by AP
are presented. The size of the filled circle indicates the AP score, while the color of the circle means the number of citations (i.e., CC score). The higher an author's CC score,
the darker the color of the circle. Two authors are connected if they have collaborated at least once. The width of the edge between two nodes indicates the number of papers that they have collaborated.}\label{network}
\end{figure*}

\begin{figure}[h!]
\centering \scalebox{0.58}[0.58]{\includegraphics{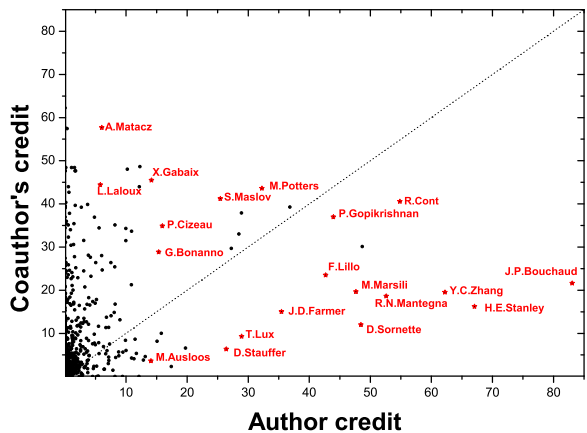}
}\caption{(Color online) The author's score as a function of the average score of all his/her co-authors'.}\label{relation}
\end{figure}

\subsection{Ranking of Scientists}
Figure ~\ref{author} shows the scatter plot of AP rank versus CC rank for authors.
If the two methods provide the same ranking, all the points would fall on the
diagonal. It shows that AP rank can provide a score that is in general proportional to CC rank. However, there are some deviations. We apply Kendall's tau ($\tau$) coefficient \cite{Kendall1938} to measure their correlation, which is  defined as
\begin{equation}
\tau=\frac{n_c-n_d}{n_t},
\end{equation}
where $n_c$ and $n_d$ are the numbers of concordant pairs and discordant pairs, respectively. $n_t$ is the total number of pairs. The rank correlation between AP rank and CC rank is 0.784. The points below the diagonal are the scientists that have higher scores by AP rank while a smaller score by CC rank,
which means that these scientists are more important in this field
than merely the number of citations indicate. For example, J. D. Farmer who owns 19
papers in our dataset has only been cited 197 times and has a CC
rank of 30, while his AP rank is 13. The reason is that most of his
citations come from prominent scientists and thus his prestige is improved.

The co-authorship network in the field of
econophysics is shown in Fig.~\ref{network} where the top-150 scientists
ranked by AP method are presented. The size of the filled circle
indicates the AP score of author. The higher an author's AP
score, the larger the circle. The color of the circle represents the
number of citations, namely author's CC score. The higher an author's CC
score, the darker the color of the circle. The width of the edge between two
nodes is proportional to the number of papers that these two authors
collaborated. From this figure, we can find a very clear community
structure. The largest community is lead by H. E. Stanley from
Boston University. The corresponding big and red circle indicates that he is a
prominent scientist in the field of econophysics.

Another interesting thing is to investigate the role of a scientist in
his/her community. Figure~\ref{relation} shows the author's score as a function of the average score of all his/her co-authors. If an author's score is larger than the average of his/her co-authors (i.e., below the diagonal), it means that his/her overall influence is more than his/her co-authors' and thus he/she may probably play a leading role in the group. On the contrary, if an author's score is much lower than the average of his/her co-authors, he/she is more likely to be a follower (e.g., a student). Therefore, this method provides a potential way to identify the supervisor-supervisee relationship.

\begin{figure}[h!]
\centering \scalebox{0.58}[0.58]{\includegraphics{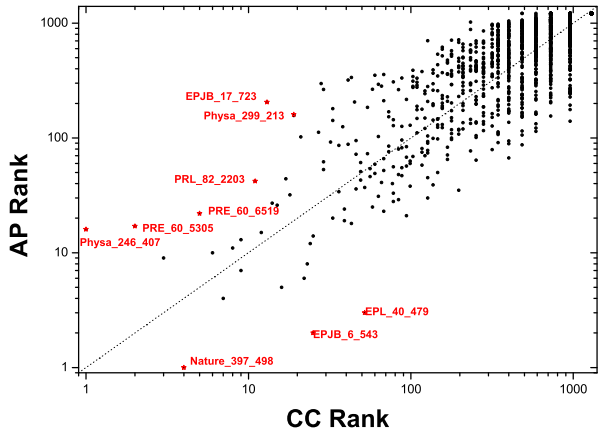}
}\caption{(Color online) Scatter plot of AP rank \emph{vs.} CC
rank for papers. If the two methods provide the same ranking all the
points would fall on the diagonal. The outliers indicate the
remarkable difference between AP rank and CC rank. The Kendall's $\tau$ coefficient is 0.644.}\label{paper}
\end{figure}

\subsection{Ranking of Papers}
\begin{figure}[h!]
\centering \scalebox{0.2}[0.2]{\includegraphics{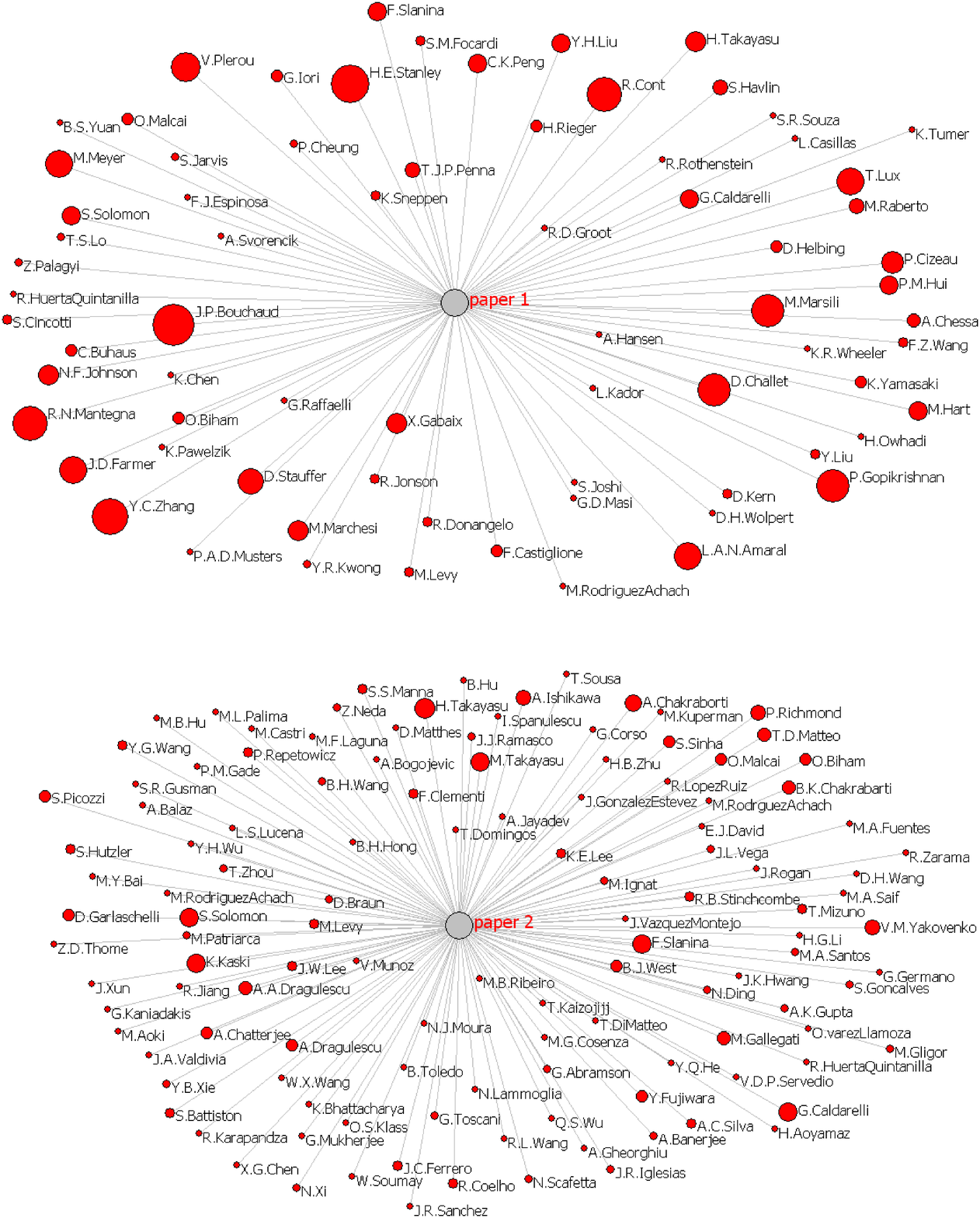}
}\caption{(Color online) Two typical examples of papers. (a) Paper 1
has 38 citations yet very high AP rank. (b) Paper 2 has 70 citations
yet very low AP rank. The size of the circle indicates the author's AP score. The higher an author's score, the larger the circle.}\label{paper_example}
\end{figure}

Figure ~\ref{paper} shows the comparison of the AP rank
and CC rank for papers. The Kendall's $\tau$ coefficient of AP rank and CC rank is 0.644. Comparing with the result on ranking authors, the difference between the results ranked by AP and CC is comparatively larger when ranking papers. Some typical outliers are labeled by stars with their publication information, including the journal, volume and starting page. For example, ``Nature\_397\_498" indicates that this paper is published in volume 397 of \emph{Nature} and starts from page 498. As AP algorithm mainly focuses on the interactions between papers and authors, the
citations of a paper from some low-score authors have small influence to
the paper's ranking result. In contrast, the citations from the
prominent scientists will contribute more to the paper's score. We
give two typical examples in Fig.~\ref{paper_example}. Although the
first paper shown in Fig.~\ref{paper_example}(a) [EPL 40 (1997) 479] has only been cited
by 38 times and has a CC rank of 52, we rank it the third since it was
cited by many high credit authors indicated by large circles. In contrast,
although the paper in Fig.~\ref{paper_example}(b) [Physica A 299 (2001) 213] has 70 citations
and has been ranked 19 by CC rank, most of these citations come from papers
written by low-score authors, it's ranked only 158 by AP.

\begin{figure}[h!]
\centering \scalebox{0.58}[0.58]{\includegraphics{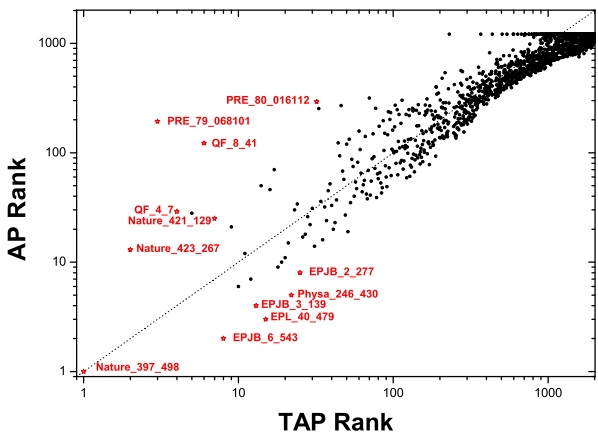}
}\caption{(Color online) Scatter plot of AP rank \emph{vs.} TAP rank for papers.}\label{time}
\end{figure}

In general, the papers published years ago are more likely to attract attentions and cumulate citations than the recently published papers. The most cited papers, which are usually considered as the representative or important works in the related field, will further receive more and more citations. As a result, old papers tend to obtain higher ranks than the fresh papers due to the cumulative effect as time goes on. We therefore define a time-dependent AP rank method (TAP), with the score of a paper $\alpha$ given by
\begin{equation}
Q_{p{\alpha}}^{\mathrm{TAP}}=\frac{Q_{p{\alpha}}^{\mathrm{AP}}}{T_0-T_{\alpha}},
\end{equation}
where $Q_{p{\alpha}}^{\mathrm{AP}}$ is the final score of paper $\alpha$ obtained by AP rank. The denominator is the number of months between the publication month of paper $\alpha$ (i.e., $T_\alpha$) and the observing month (i.e., $T_0$). For our dataset, $T_0$ is September 2010. Figure~\ref{time} shows the scatter plot of AP rank versus TAP rank for papers. The papers of both high AP and high TAP rank are usually the prominent works and have long-term influence in the related research field. One typical example is the paper ``Scaling and criticality in a stochastic multi-agent model of a financial market" written by T. Lux and M. Marchesi in 1999. This paper is the first of both AP rank and TAP rank. Moveover, some works with very high AP rank while a relative low TAP rank are usually the pioneer works that published many years ago with overall high influence, such as the papers ``EPJB 6 (1998) 543", ``EPL 40 (1997) 479", ``EPJB 3 (1998) 139" and ``Physica A 246 (1997) 430", \emph{etc}. In contrast, the papers with not very low AP rank but very high TAP rank (see the outliers in the top left corner of Fig.~\ref{time}) are usually recently published papers and potentially high influence papers in the future. Typical examples are ``PRE 80 (2009) 016112", ``PRE 79 (2009) 068101" and ``Quantitative Finance 8 (2008) 41". In addition, some works which already have a high AP rank, are ranked even higher when publication date is considered. For example, ``Nature 423 (2003) 267", ``Nature 421 (2003) 129" and ``Quantitative Finance 4 (2004) 7" are respectively ranked 14, 26 and 29 by AP method, indicating their high influence in the field of econophysics, and their corresponding TAP ranks are 2, 7 and 4. These works are very promising and may become more influential in the future.

\subsection{Evaluation of Journals}

\begin{figure}[h!]
\centering \scalebox{0.78}[0.78]{\includegraphics{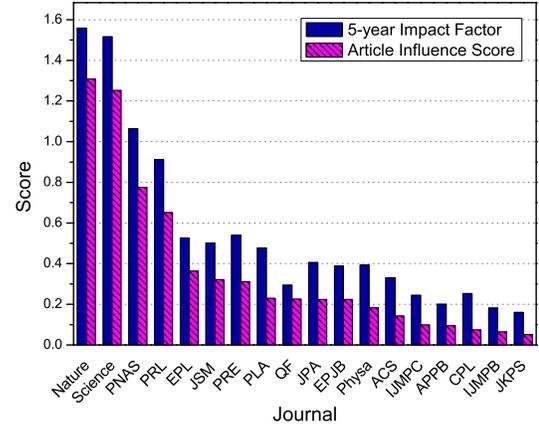}
}\caption{(Color online) The 5-year Impact Factors and Article Influence Scores of eighteen journals in our dataset.
For better presentation, the scores are modified in the form $\mathrm{log}_{10}(s+1)$, where $s$ is the real value of 5-year Impact Factor or Article Influence Score.}\label{JIF}
\end{figure}

\begin{figure}[h!]
\centering \scalebox{0.78}[0.78]{
\includegraphics{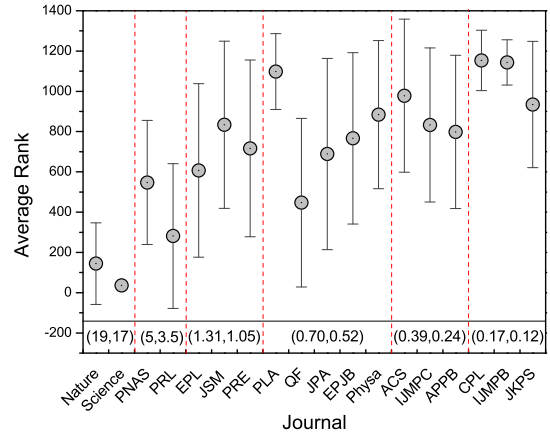}
}\caption{The average rank of the papers in each journal.
The journals are ranked from left to right in descending order by their Article Influence Scores. At the bottom of each region, the left number and the right number indicate the largest and the smallest Article Influence Scores of the journals in this region, respectively.}\label{aveRank}
\end{figure}

\begin{figure}[h!]
\centering \scalebox{0.78}[0.78]{
\includegraphics{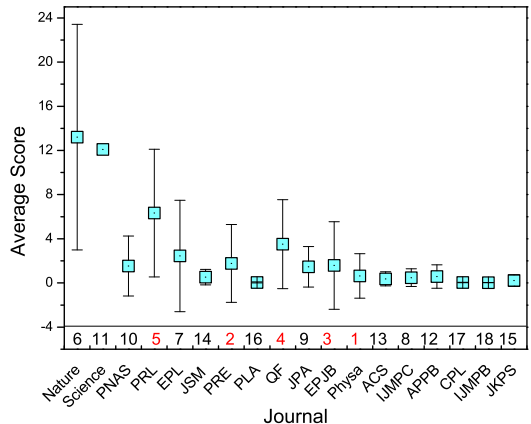}
}\caption{The average score of the papers in each journal.
The journals are ranked from left to right in descending order by their Article Influence Scores. The numbers at the bottom indicate the ranks of journals obtained by overall influence score.}\label{aveScore}
\end{figure}

\begin{table}
\caption{Mainstream journals in econophysics ranked by their overall influence scores. Their corresponding information of AIS, 5-year IF, the number of publications, average score (AvgS) and overall influence score (OIS) are also presented.\label{RankJ}}
\centering
{\begin{tabular}{cl ccccc} \hline
Rank &Journal & AIS &5-year IF &Paper &AvgS &OIS\\
\hline
1	&Physa	&0.522	&1.467	&1120	&0.640	&716.845\\
2	&PRE 	&1.047	&2.458	&179	&1.772	&317.269\\
3	&EPJB	&0.674	&1.443	&148	&1.585	&234.509\\
4	&QF	&0.682	&0.968	&52	&3.513	&182.669\\
5	&PRL	&3.486	&7.154	&20	&6.333	&126.651\\
6	&Nature	&19.334	&35.241	&5	&13.205	&66.026\\
7	&EPL	&1.308	&2.358	&21	&2.448	&51.410\\
8	&IJMPC	&0.256	&0.753	&47	&0.4759	&22.365\\
9	&JPA	&0.675	&1.542	&14	&1.456	&20.378\\
10	&PNAS	&4.959	&10.591	&12	&1.534 &18.405\\
11	&Science	&16.859	&31.769	&1	&12.083	&12.082\\
12	&APPB	&0.243	&0.586	&11	&0.577	&6.345\\
13	&ACS	&0.39	&1.141	&15	&0.371	&5.561\\
14	&JSM 	&1.094	&2.169	&10	&0.527	&5.269\\
15	&JKPS	&0.124	&0.446	&18	&0.213	&3.829\\
16	&PLA	&0.697	&1.995	&31	&0.046	&1.440\\
17	&CPL	&0.186	&0.79	&7	&0.035	&0.246\\
18	&IJMPB	&0.159	&0.519	&7	&0.031	&0.216\\
\hline
\end{tabular}}
\end{table}

We further investigate the correlation between quality of the papers and the quality of their corresponding published journals. As we know, the \emph{ISI Impact Factor} (IF) which is defined as the mean number of citations a journal receives over a 2 year period is widely used to evaluate the quality, importance or influence of journals. However, we argue that since IF is based solely on the number of citations regardless of the prestige of citing sources, it can only be considered as a metric of popularity, and thus is inappropriate to be used to quantify the quality or prestige of journals \cite{Bollen2006}. Therefore we consider another metric, the \emph{Article Influence$^{TM}$ Score} (AIS) \cite{Bergstrom2007}, as an indicator to reflect the quality of a journal. Unlike IF, AIS weights each citation by the quality of the citing journals. In Fig.~\ref{JIF}, we compare the 5-year IF and AIS of eighteen selected journals, including \emph{Nature}, \emph{Science} and the journals listed in Table ~\ref{Journal}. The data, shown in Table ~\ref{RankJ}, is obtained from the Thompson Reuters' \emph{2010 Journal Citation Report} (JCR). IJTAF is excluded since it does not have record in JCR. All the journals are ranked in descending order according to their AIS. For most journals, the 5-year IF and the AIS are positively correlated, but there are a few exceptions. PRE has higher 5-year IF than EPL while with lower AIS. Although QF has a low 5-year IF, its AIS is even higher than some journals with larger 5-year IF. On the contrary, CPL has high 5-year IF while with very low AIS. More detailed comparison and discussion of these two measures can be found in Ref. \cite{Rizkallah2010}.

Figure~\ref{aveRank} shows the average rank of the papers in each journal. The eighteen selected journals are ranked in descending order by their AIS. Generally speaking, papers published in high AIS journals tend to be ranked higher than that published in low AIS journals. The average score of papers in each journal is presented in Fig.~\ref{aveScore}. It shows that the average score of papers in high AIS journal is likely to be higher than in low AIS journal. Interestingly, we find that the journal QF has even lower average rank and higher average score than some high AIS journals, such as PRE, EPL, etc. This may indicate that ``Quantitative Finance" is a mainstream journal of econophysics, and thus its papers are likely to have larger influence in this field. Finally, we use the product of average score and the number of published papers to quantify the overall influence of a journal in econophysics, according to which we can find the mainstream journals in this field. The result is shown in Table ~\ref{RankJ}, where the journals are ranked by their overall influence scores (OIS) and their corresponding information of AIS, 5-year IF, the number of publications and average scores (AvgS) are also presented. As we can see, the top-5 mainstream journals in econophysics are ``Physica A", ``Phys Rev E", ``Eur Phys J B", ``Quant Financ" and ``Phys Rev Lett".

\section{Discussion}

In this paper, we proposed an iterative algorithm named \emph{AP Rank} to quantify the scientists' prestige and the quality of their publications via their inter-relationship on an author-paper bipartite network. The rationale behind this method is that a paper is expected to be of high quality if it was cited by prestigious scientists, while high quality papers will in turn raise its authors' prestige. It is thus clear that AP rank weighs the prestige of quoters more than the number of citations.
The former is referred to the prestige while the latter to the popularity. We conducted the experiment on the dataset consists of 1990 scientists and their 2012 papers in econophysics, and compared the ranking results with the citation counts. Although these two methods have overlap to some extent (for authors, Kendall's $\tau=0.784$, for papers, Kendall's $\tau=0.644$), the outliers reveal the remarkable and meaningful differences. We found that some scientists with lower CC rank may have higher influence than that indicated by their citations, because they are appreciated by prestigious scientists. Some papers with large number of citations are ranked lower by AP rank, indicating that they are over evaluated by merely counting the number of citations. In other words, these papers are popular, but not prestigious. The fact that a paper can only cite earlier papers makes the publishing time an important factor in the paper citation network. Therefore, the old papers will have larger opportunity to accumulate more citations than fresh works. With this consideration, we proposed a time-dependent AP rank (TAP rank). The papers can be classified by synthetically considering their AP rank and TAP rank. We further evaluated the influence of journals by the total ranking score of its publications. Top-5 mainstream journals in econophysics were founded: ``Physica A", ``Phys Rev E", ``Eur Phys J B", ``Quant Financ" and ``Phys Rev Lett". In reality, our method can be directly applied to quantify the journals' quality by constructing a journal-paper bipartite network where the citations between journals are considered.

The main advantages of AP rank are obvious: i) it is parameter-free; ii) it considers the interaction between the prestige of scientists and the quality of their publications; iii) it is effective to distinguish the prestige from popularity. In addition, like PageRank algorithm or its variants for ranking task in other systems ranging from social webs \cite{LeaderRank2011} to ecosystem \cite{Allesina2009}, the AP rank method can also be generalized to applications in a wide range of systems. The modifications and extensions of this method are easy to be implemented. Take the micro-blog web (e.g., Twitter, Sina, etc.) as an example, under the framework of AP rank we can build an online reputation system to identify the influential users and evaluate the quality of their blogs (e.g., tweets) via constructing a bipartite network where the forwarding relation between micro-blogs (e.g., retweet) can be considered as a kind of citation.

How to well utilize the available information to devise a good evaluation or ranking method has been a long-lasting challenge. As an issue of ranking metrics, problems arise: Is it simple to calculate? Dose it reflect the intrinsic value? Is it robust against manipulations? Since every indictor will have its own strengths and weaknesses, it is difficult to design a panacea-like metric that covers all aspects. For instance, citation counts is very simple but not robust against manipulations. The rank can be easily increased by the abuse of self-citations or cross-citations within a small group. This is the shortcoming of all citation-based metrics, including the \emph{h} index and the Impact Factor of journals. In addition, how to make comparison in different scientific fields is also important when designing a metric. Some progresses have been made in this direction \cite{Xie2007,Petersen2010}. For sure in the near future, with the advance of technology, more information and data can be conveniently obtained, and are expected to foster the design of better ranking metrics to face these challenges.

\acknowledgments
We thank Yougui Wang, Chi Ho Yeung, Yi-Cheng Zhang, Changsong Zhou and Tao Zhou for helpful suggestions. This work is supported by the Swiss National Science Foundation under Grant No. 200020-132253. M.L. is supported by the DSTA of Singapore under Project Agreement No. POD0613356.

\end{document}